\documentclass[aps,prc,showpacs,twocolumn,preprintnumbers,floatfix,
showkeys,tightenlines]{revtex4}
\usepackage{amsmath,amssymb,amsfonts}
\usepackage{graphicx}
\usepackage{color}

\allowdisplaybreaks
\newcommand{\vece}{\vec{\mathcal{E}}}
\newcommand{\veces}{\vec{\tilde{\mathcal{E}}}}
\newcommand{\vecet}{\vec{\mathcal{E}_t}}
\newcommand{\vecf}{\vec{\mathcal{F}}}
\newcommand{\ce}{\tilde{\mathcal{E}}}
\newcommand{\vecb}{\vec{\mathcal{B}}}
\newcommand{\vecr}{\vec{r}}
\newcommand{\vecv}{\vec{v}}
\newcommand{\veck}{\vec{k}}
\begin{document}
\title{Energy gain of heavy
quarks by fluctuations in the QGP}
\author{Purnendu \surname{Chakraborty}}
\email{purnendu.chakraborty@saha.ac.in}
\author{Munshi Golam \surname{Mustafa}}
\email{musnhigolam.mustafa@saha.ac.in}
\affiliation{Theory Division, Saha Institute of Nuclear Physics, 1/AF
  Bidhannagar, Kolkata 700064.}
\author{Markus H. \surname{Thoma}}
\email{mthoma@ipp.mpg.de}
\affiliation{Max-Planck-Institut
f\"ur extraterrestrische Physik, P.O. Box 1312, 85741 Garching, Germany} 
\begin{abstract}
The collisional energy gain of a heavy quark due to chromo-electromagnetic 
field fluctuations in a quark-gluon plasma is investigated. 
The field fluctuations lead to an energy gain of the quark for all 
velocities. The net effect is a reduction of the collisional 
energy loss by 15-40\% for parameters relevant at RHIC energies.
  
\end{abstract}
\preprint{SINP/TNP/06-30}
\maketitle

The aim of the ongoing relativistic heavy-ion collision experiments is to 
explore the possible plasma phase of QCD, the so called quark-gluon plasma 
(QGP). High energy partons produced in initial partonic sub-processes in 
collisions between two heavy nuclei will lose their energy while propagating 
through the dense matter formed after 
such collisions, resulting in jet quenching. The amount of quenching depends 
upon the state of the fireball produced and the resulting quenching pattern 
may be used for identifying and investigating the plasma 
phase~\cite{Gyulassy:1990ye}. 
In order to quantitatively understand medium modifications of hard parton
characteristics  
in the final state, the energy loss of partons in the QGP has to be 
determined. There are two contributions to the energy loss of a parton 
in a QGP: one is caused by elastic collisions between the partons and the 
other is caused 
by radiative losses. It is generally believed that the radiative loss 
due to multiple gluon 
radiation (see~\cite{Baier:2000mf,Gyulassy:2003mc} for a review) 
dominates over the collisional one in the ultra-relativistic case. 
However, it has been shown recently 
that for realistic values of the parameters relevant for heavy-ion collisions, 
there is a wide range of parton energies in which the 
magnitude of the collisional loss is comparable to the radiative loss for 
heavy~\cite{Mustafa:2003vh,Mustafa:2004dr,Wicks:2005gt,Moore:2004tg} 
as well as 
for light~\cite{Mustafa:2003vh,Roy:2005am} quark flavors. 
           
Earlier estimates of the collisional energy loss in the QGP~\cite{Thoma:1990fm,
Mrowczynski:1991da}
were obtained by treating the medium in 
an average manner, {\it i.e.}, microscopic fluctuations were  neglected. 
However, the QGP, being a statistical system, is characterized by 
omnipresent stochastic fluctuations.  These fluctuations couple to external 
perturbations and the response of the medium to these perturbations can be 
expressed through suitable correlation functions of the microscopically 
fluctuating variables.  
Now, it is well known that the motion of charged particles in 
such an environment 
is stochastic in nature and resembles Brownian motion. The Fokker-Planck 
equation provides a natural basis for a differential characterization of 
such a stochastic motion.  
In a homogeneous and isotropic plasma, the Fokker-Planck equation can be
recast in the form 
\begin{equation}
\label{fokker-planck}
\frac{\partial f}{\partial t} = \sum_{n=1}^{\infty} 
 \frac{\left(-1\right)^n}{n!} \frac{\partial^n}{\partial p_i \partial p_j \cdots
\partial p_n}\left[{\mathcal D}_{ij\cdots n}\left(p,t\right) 
 f\left(p,t\right)\right]\,,
\end{equation} 
where $f\left(p,t\right)$ is the phase-space distribution of the test particle.
$D_i$ in~(\ref{fokker-planck}) is the drag coefficient - a quantity 
closely related to energy loss per unit length 
$dE/dx$~\cite{Mustafa:2004dr,
Moore:2004tg,Svetitsky:1987gq,Pal:1997,Walton:1999dy}. 
It is to be noted that starting from $n=2$, all terms in~(\ref{fokker-planck})
are statistical  in origin, {\it i.e}, they arise due to microscopic field 
fluctuations, to leading order in $\alpha_s$~\cite{gasirowicz,Hubbard}.
For the drag coefficient or energy loss, the fluctuating field as well as 
the polarization field contribute as will be seen later.  
The effect of field fluctuations on the passage of a 
charged particle 
through a non-relativistic classical plasma has been worked out by 
several 
authors \cite{Hubbard,sitenko1,sitenko2,kalman,gasirowicz,thompson,ichimaru} 
in the past. 
This effect leads to an energy gain of the particles and is most effective 
in the low velocity limit. 
Given the fact that the subject of the energy loss is of topical interest, 
 it is 
the principal motivation 
of the present article to quantitatively estimate the effect of microscopic 
 electromagnetic 
 fluctuations on the energy loss of a heavy quark passing through an 
equilibrium, weakly-coupled QGP within the semiclassical approximation. 
 
The semiclassical approach was adopted earlier to calculate 
the collisional energy loss of a heavy quark due to polarization effects 
of the medium~\cite{Thoma:1990fm,Mrowczynski:1991da}.  
It is assumed that the energy 
lost by the particle  
per unit time is small compared to the energy of the particle 
itself so that the change in the velocity of the particle during the 
motion may be neglected, {\it i.e}, the particle moves in a straight line 
trajectory. The energy loss of a particle is determined by the work of the 
retarding forces acting on the particle in the plasma from the chromo-electric 
field generated by the particle itself while moving. So the energy loss of the 
particle per unit time is given by \cite{Thoma:1990fm},
{\begin{equation}
\label{dedt}
\frac{dE}{dt} = Q^a\vecv\cdot\vece^a|_{\vecr=\vecv t}\,,
\end{equation}    
where the field is taken at the location of the particle.  
Within linear response theory the correlation function of the fluctuations 
of charge and current densities and the electromagnetic fields in the 
medium with space-time dispersion are completely determined in terms of
the dielectric tensor of the medium.
In the Abelian approximation, the total chromo-electric field 
$\vec{\mathcal{E}}^a$ induced  in the QGP can 
be related to the external current of the test charge by solving  
Maxwell's equations and the equation of continuity
\begin{equation}
\label{e.and.j}
\left[\epsilon_{ij}\left(\omega,k\right) - \frac{k^2}{\omega^2}\left
(\delta_{ij} - \frac{k_ik_j}{k^2}\right)\right ]
\mathcal{E}^a_j\left(\omega,k\right) 
= \frac{4\pi}{i\omega}j^a_i\left(\omega,k\right) ,
\end{equation}       
with the color charge current $\vec j^a$. It should be noted that in this 
approximation, the dielectric tensor $\epsilon_{ij}$ is diagonal in 
the color indices~\cite{Mrowczynski:1989bv}. 
In an isotropic and homogeneous plasma the dielectric tensor $\epsilon_{ij}$ 
can be decomposed into longitudinal and transverse parts, 
\begin{equation}
\label{epsilon_ij}
\epsilon_{ij}\left(\omega,k\right) = \epsilon_L\left(\omega,k\right)
\mathcal{P}_{ij}^L
+ \epsilon_T\left(\omega,k\right)\mathcal{P}_{ij}^T\,,
\end{equation}
where, \(\mathcal{P}_{ij}^L = k_i k_j/k^2\) and \(\mathcal{P}_{ij}^T = 
\delta_{ij} - k_i k_j/k^2\). 
The gauge invariant high-temperature expression for the dielectric functions
which is in accordance with the Abelian approximation 
(see below) are given by 
(see e.g. \cite{Mrowczynski:1989bv,thoma1995}) 
\begin{eqnarray}
\epsilon_L(\omega,k)&=& 1+\frac{m^2_D}{k^2} \left [ 1- \frac{\omega}{2k} 
\left ( \ln \left |\frac{\omega+k}{\omega-k}\right | \right.\right.\nonumber\\
&&\left.\left.-i\pi \Theta(k^2-\omega^2) \right ) \right ] 
\, \ , \nonumber \\
\epsilon_T(\omega,k) &=& 1- \frac{m^2_D}{2\omega^2}
\left [ \frac{\omega^2}{k^2}
 + \left (1-\frac{\omega^2}{k^2} \right )
\frac{\omega}{2k} \left ( \ln \left | 
\frac{\omega+k}{\omega-k}\right | \right.\right.\nonumber\\
&&\left.\left. -i\pi \Theta(k^2-\omega^2)  \right)\right] 
\, \, , \label{dielectric_lt}
\end{eqnarray}
where $m_D^2 = g^2T^2\left(1 + N_f/6\right)$ is the Debye mass
squared. Substitution of (\ref{e.and.j}) together with  (\ref{epsilon_ij}) 
and  (\ref{dielectric_lt}) in (\ref{dedt}) gives the polarization loss 
of the moving parton~\cite{Thoma:1990fm,Mrowczynski:1991da}. 

The previous formula for the energy loss in (\ref{dedt}) does not take 
into account the field fluctuation in the plasma and the particle recoil in
collisions.
To accommodate these effects it is necessary to 
replace (\ref{dedt}) with \cite{sitenko1,sitenko2},
{\begin{equation}
\label{dedt_stat}
\frac{dE}{dt} =\left \langle Q^a \vecv\left(t\right)\cdot\vece^a_t\left(\vecr\left(t\right),
t\right)\right \rangle\,,
\end{equation}
where
$\left\langle\cdots\right\rangle$ denotes the statistical
averaging operation. It is to be noted that here two averaging procedures are performed: 
I) an ensemble average {\it w.r.t} the equilibrium density matrix and 
II) a time average over random fluctuations in plasma. 
These two operations are commuting and only after both of them are 
performed the average quantity takes up a smooth value \cite{kalman}.
In the following, we will explicitly denote the ensemble average by  
$\left\langle \cdots \right\rangle_\beta$ wherever required to avoid possible 
confusion.  

The electric field $\vecet$ in (\ref{dedt_stat}) consists of the induced field 
 $\vece$ given by (\ref{e.and.j}) and a spontaneously generated microscopic 
field $\veces$, the latter being a random function of position and time. 

The classical equation of motion of the  particles 
in the electromagnetic field have the form,
\begin{equation}
\frac{d\vec{p}}{dt} = Q^a \left[\vece^a_t\left(\vecr
\left(t\right), t\right) + \vec{v}\times \vecb^a_t\left(\vecr
\left(t\right),t\right)\right]
\,.\label{eqmotion1}
\end{equation}

Integrating the equation of motion  
(\ref{eqmotion1}) we find, 
\begin{eqnarray}
\vecv\left(t\right) &=& \vecv_0 + \frac{1}{E_0}\int_0^t\,dt_1\,Q^a
 \vecf^a_t\left(\vecr\left(t_1\right), t_1\right)\,,\nonumber\\  
\vecr\left(t\right) &=& \vecv_0 t+ 
\frac{1}{E_0}\int_0^t\,dt_1\int_0^{t_1}dt_2\,Q^a 
\vecf^a_t\left(\vecr\left(t_2\right), t_2\right)\,.
\label{eqmotion2}
\end{eqnarray}
where $\vecf_t = \vecet + \vecv \times \vecb_t$ and $E_0$ is the 
initial parton energy. The magnetic field  $\vecb_t$ consists
of an induced and a microscopic part as in the case of $\vecet$.
The mean change in the energy of the particle per unit time is given by 
(\ref{dedt_stat}). The stochastic time dependence as embodied 
in (\ref{dedt_stat}) comes because of explicit fluctuations in time and motion 
of the particle from one field point to another. The dependence on the latter 
will be expanded about the mean rectilinear motion where the former 
will be left untouched. 
This is done by picking a time interval 
$\Delta t$ sufficiently large with respect 
to the time scale of random electromagnetic fluctuations in the plasma $(\tau_2)$
but small compared with the time during which the particle motion changes 
appreciably $(\tau_1)$, 
\begin{equation}
\label{scale}
\tau_1 \gg \Delta t \gg \tau_2\,.
\end{equation}
Parametrically, 
$\tau_1 \sim \left[g^4T \ln{(1/g)}\right]^{-1}$ and $\tau_2 \sim 
\left(gT\right)^{-1}$~\cite{Arnold:2002zm}. 
In the weak coupling limit, these scales are widely 
separated.     
Keeping only the leading order terms in the expansion we 
get (see eq. 8.31 in \cite{sitenko1} and eq. 13.1.3.4 in \cite{sitenko2}) 
\begin{eqnarray}
&&\vecv\left(t\right) = \vecv_0 + \frac{1}{E_0}
\int_0^t\,dt_1 Q^a
 \vecf^a_t\left(\vecr_0\left(t_1\right), t_1\right)\,,\nonumber\\
&&\vece^a_t\left(\vecr\left(t\right), t\right) 
= \vece^a_t\left(\vecr_0\left(t\right), t\right) + 
 \frac{Q^b}{E_0}\int_0^t
\,dt_1\int_0^{t_1}\,dt_2\, \nonumber\\
&& \, \,  \times \sum_j \mathcal{E}_{j,t}^b\left(\vecr_0\left(t_2
\right), t_2\right)\,\frac{\partial}{\partial r_{0j}}\,
\vece^a_t\left(\vecr_0\left(t\right), t\right)\,,\label{eqmotion4}
\end{eqnarray}    
where, $\vecr_0\left(t\right) = \vecv_0 t$.
Substituting (\ref{eqmotion4}) in 
(\ref{dedt_stat}) and 
using 
$\left \langle
\mathcal{E}_i^a \mathcal{B}_j^a\right\rangle_\beta=0$~\cite{landaustaphys}, 
we get       
\begin{eqnarray}
\frac{dE}{dt} &=& \left\langle Q^a\,\vecv_0\cdot\vece^a_t
\left(\vecr_0
\left(t\right), t\right)\right\rangle_\beta \nonumber\\
&+& \frac{Q^a Q^b}{E_0}\int_0^t\,dt_1
\left\langle\vece^b_t\left(\vecr_0\left(t_1\right), t_1\right)\cdot
\vece^a_t\left(\vecr_0\left(t\right), t\right)\right\rangle_\beta\nonumber \\
&+&  \frac{Q^a Q^b}{E_0}\int_0^t\,dt_1\int_0^{t_1}dt_2\,\left\langle\sum_j
\mathcal{E}_{t,j}^b\left(\vecr_0\left(t_2\right), t_2\right)\right.
\nonumber\\
&\times&\left.\frac{\partial}
{\partial r_{0j}}\,\,\vecv_0\cdot\vece^a_t
\left(\vecr_0\left(t\right), t\right)\right\rangle_\beta\,. \label{eqmotion5}
\end{eqnarray}
Since the mean value of the fluctuating part of the field equals zero, 
$\left\langle \veces \right\rangle_\beta = 0$, 
$\left\langle
\vece_t\left(\vecr\left(t\right),t\right)\right\rangle_\beta$ 
equals the chromo-electric field 
produced by the 
particle itself in the plasma. The first term in (\ref{eqmotion5}) 
therefore 
corresponds to the usual polarization loss of the parton calculated in 
\cite{Thoma:1990fm}. Provided there exists a hierarchy 
of scales (\ref{scale}), 
 it can be shown that the polarization field 
does not contribute to leading order in the correlations functions appearing 
in the second and third terms in 
 (\ref{eqmotion5}) as in the case of higher order Fokker-Planck 
coefficients~\cite{Hubbard,ichimaru}
. These terms  
correspond to the statistical change in the energy of the moving parton in 
the plasma 
due to the fluctuations of the chromo-electromagnetic fields as well as 
the velocity of the particle under the influence of this field. 
The second term in (\ref{eqmotion5}) 
corresponds to the statistical part of the dynamic friction 
due to the space-time correlation in the fluctuations in the electrical field
whereas the third one 
corresponds to the average change in 
the energy of the moving particle due to the correlation between the 
fluctuation in 
the velocity of the particle and the fluctuation in the electrical field
in the plasma.  
The temporal averaging in  (\ref{eqmotion5}), by definition, 
includes many random fluctuations over the mean 
motion. However,  
the correlation function of these fluctuations are suppressed 
 beyond their characteristic time scales. 
This allows us to formally extend the upper limits in 
time-integrations in (\ref{eqmotion5}) to infinity. 

Now within linear response theory,  
 the power spectrum of the chromo-electromagnetic fields follows
from the fluctuation-dissipation theorem and is completely determined by the
dielectric functions of the medium \cite{sitenko1,sitenko2},
\begin{equation}
\left\langle \tilde{\mathcal{E}}_i^a \tilde{\mathcal{E}}_j^b 
\right\rangle_{\beta;\,\omega,k} = 
\left\langle \tilde{\mathcal{E}}_i \tilde{\mathcal{E}}_j 
\right\rangle_{\beta;\,\omega,k} \delta_{ab}\,\label{eflc}
\end{equation}
where, 
\begin{equation}
\left\langle \tilde{\mathcal{E}}_i \tilde{\mathcal{E}}_j 
\right\rangle_{\beta;\,\omega,k} =
\frac{8\pi}{e^{\beta\omega}-1}\left\{
\mathcal{P}_{ij}^L \frac{{\rm Im}\,\epsilon_L}{|\epsilon_L|^2} +
\mathcal{P}_{ij}^T
\frac{{\rm Im}\,\epsilon_T}{|\epsilon_T - \eta^2|^2}
\right\} 
\label{e.and.b-fluc}
\end{equation}
with $\eta = k/\omega$.

Using the Fourier transform of  $\tilde{\mathcal{E}}_i$ together with (\ref{eflc}) 
and~(\ref{e.and.b-fluc}), we obtain from  
(\ref{eqmotion5}) the energy loss 
of the parton due to fluctuations as (see appendix),  
\begin{equation}
\left.\frac{dE}{dt}\right|_{\rm fl} = \frac{C_F\alpha_s}{16\pi^3 E}
\int\,d^3k\,\left[
\frac{\partial}{\partial\omega} 
\left\langle\omega \veces_L^2\right\rangle_\beta + 
\left\langle\veces_T^2 
\right\rangle_\beta \right]_{\omega = \veck
\cdot\vecv}\,,\label{dedt_fl1}
\end{equation}
where $\left\langle \veces_L^2\right\rangle$ and 
$\left\langle \veces_T^2\right\rangle$ denote the longitudinal and 
transverse field fluctuations, respectively, and $E = E_0$ 
is the initial energy of the parton. 

 Eq. (\ref{dedt_fl1}) can be recast as (see appendix),
\begin{eqnarray}
\left.\frac{dE}{dt}\right|_{\rm fl} &=& 
\frac{C_F\alpha_s}{8\pi^2 Ev^3}\int_0^{k_{\rm max}v}d\omega\,
\coth{\frac{\beta\omega}{2}}F\left(
\omega,k=\omega /v\right) \nonumber\\
&+& 
\frac{C_F\alpha_s}{8\pi^2 Ev}\int_0^{k_{\rm max}}dk\,k\int_0^{kv}
d\omega\,\coth{\frac{\beta\omega}{2}}G\left(\omega,k\right)\,, \nonumber\\
\label{elossfin1}
\end{eqnarray} 
where \(F\left(\omega,k\right) = 8\pi\omega^2 {\rm Im}\,
\epsilon_L/|\epsilon_L|^2\) and \(G\left(\omega,k\right) = 16\pi {\rm Im}\,
\epsilon_T/|\epsilon_T - k^2/\omega^2|^2\) and $v_0 =v$. This result is 
obviously gauge invariant if we use there the semiclassical, gauge invariant
expression for the dielectric functions (\ref{dielectric_lt}).
 
The above expression gives the mean energy (per unit time) absorbed 
by a propagating particle from the heat bath. Physically, this arises
from gluon absorption.
Thermal absorption of gluons was also shown to reduce the radiative
energy loss~\cite{Wang:2001cs}. We arrive at a somewhat similar
conclusion as there, albeit in a different context.    
It is to be noted here that since the spectral density of field fluctuations 
$\left\langle \veces^2_{L/T}\right\rangle$ are positive for 
positive frequencies by definition, according 
to (\ref{elossfin1}) the particle energy will grow due to interactions 
with the fluctuating fields. 

\begin{figure}[!htbp]
\vspace{.8cm}
\begin{center}
  \includegraphics[width=\columnwidth,keepaspectratio]{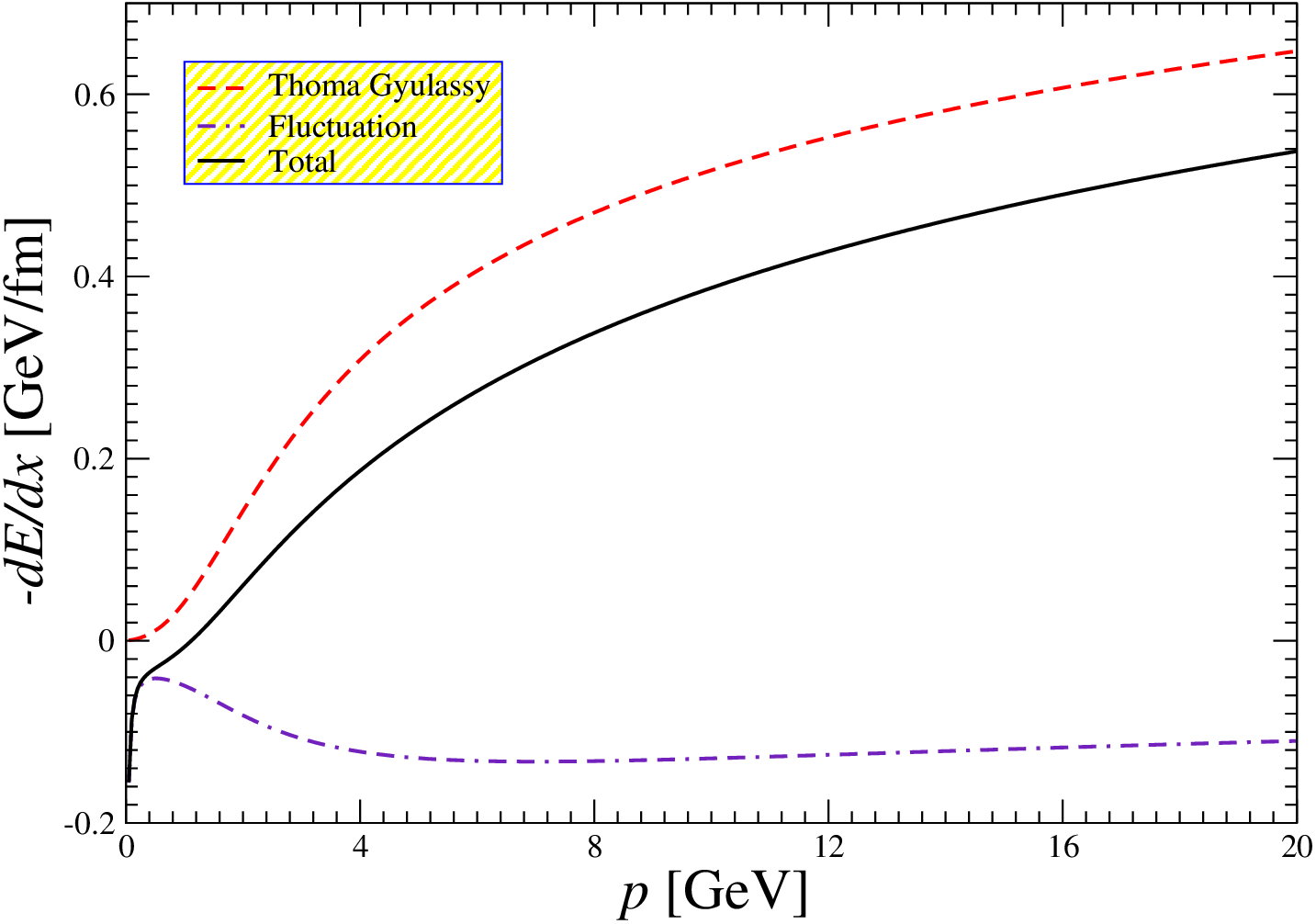}
\caption{\label{fig1} (Color online) Various contributions to the energy 
loss of charm quark in the QGP. The collisional energy loss (dashed line) 
is taken from Ref.\cite{Thoma:1990fm}.} 
\end{center}  
\end{figure}

\begin{figure}[!htbp]
\begin{center}
  \includegraphics[width=\columnwidth,keepaspectratio]{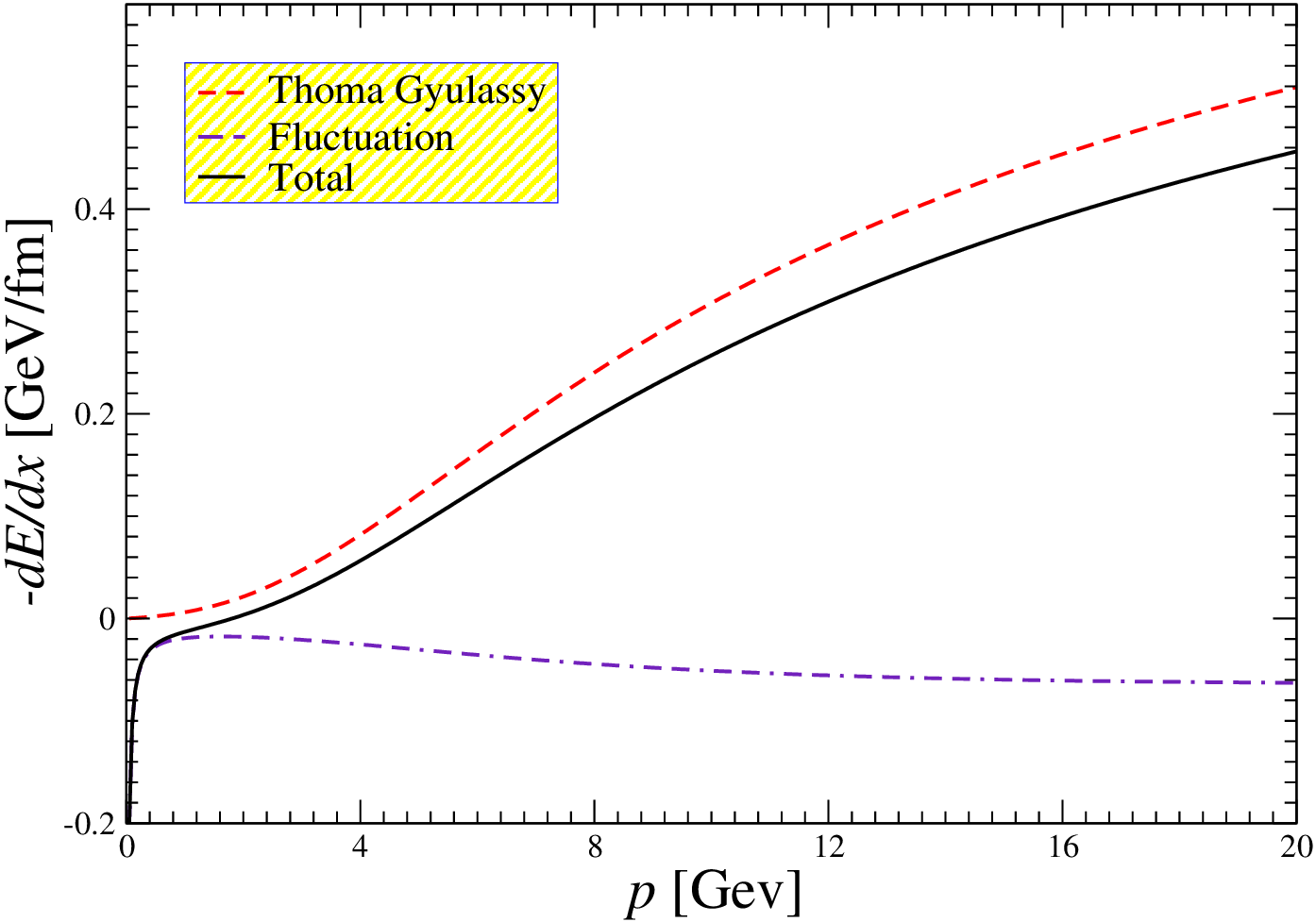}
\caption{\label{fig2} (Color online) Same as Fig.~\ref{fig1} but for a bottom 
quark.} 
\end{center} 
\vspace{.3cm} 
\end{figure}

The contribution from field fluctuations to the heavy
quark energy loss is shown in Fig.~\ref{fig1} and Fig.~\ref{fig2}. 
Our choice of the parameters is $N_f=2$, $T=250$ MeV, $\alpha_s=0.3$,
and for the charm and bottom quark masses we take 
$1.25$ and $4.2$ GeV, respectively. For 
the upper integration limit $k_{\rm max}$ we take \cite{Adil:2006ei},
\begin{equation}
\label{cutoff}
k_{\rm max} =\mbox{min}\,\left\{E,\frac{2q\left(E+p\right)}
{\sqrt{m^2+2q\left(E+p\right)}}\right\}\,,
\end{equation}
where $q \sim T$ is the typical momentum of the thermal partons of the QGP. 
     
\begin{figure}[!htbp]
\begin{center}
\includegraphics[width=\columnwidth,keepaspectratio]{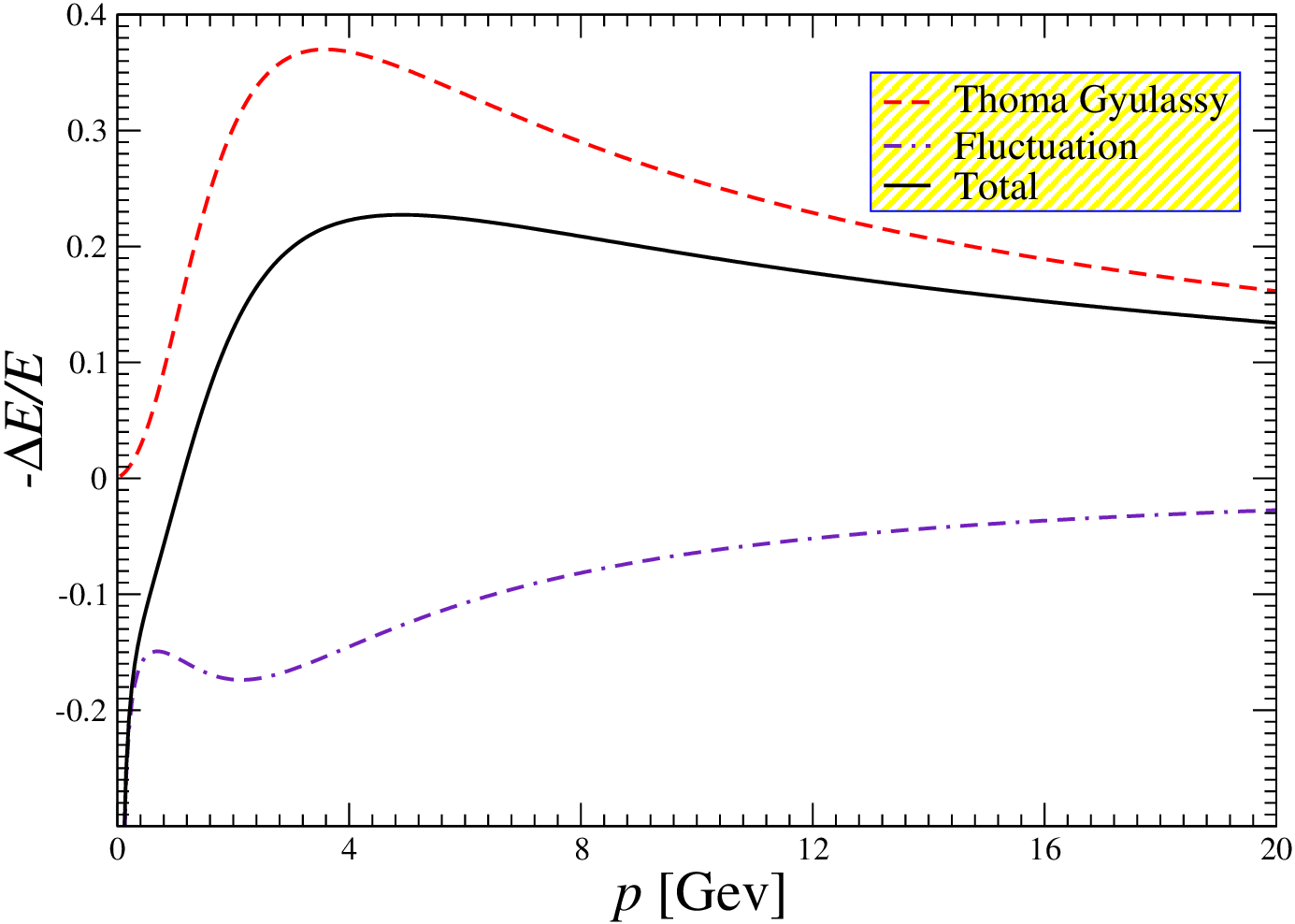}
\caption{\label{fig3} (Color online) Relative importance of the 
fluctuation loss compared to the collisional energy loss of 
Ref.\cite{Thoma:1990fm}. We take the path length to be $L=5$ fm.} 
\end{center}  
\end{figure} 

\begin{figure}[!htbp]
\vspace{.5cm}
\begin{center}
\includegraphics[width=\columnwidth,keepaspectratio]{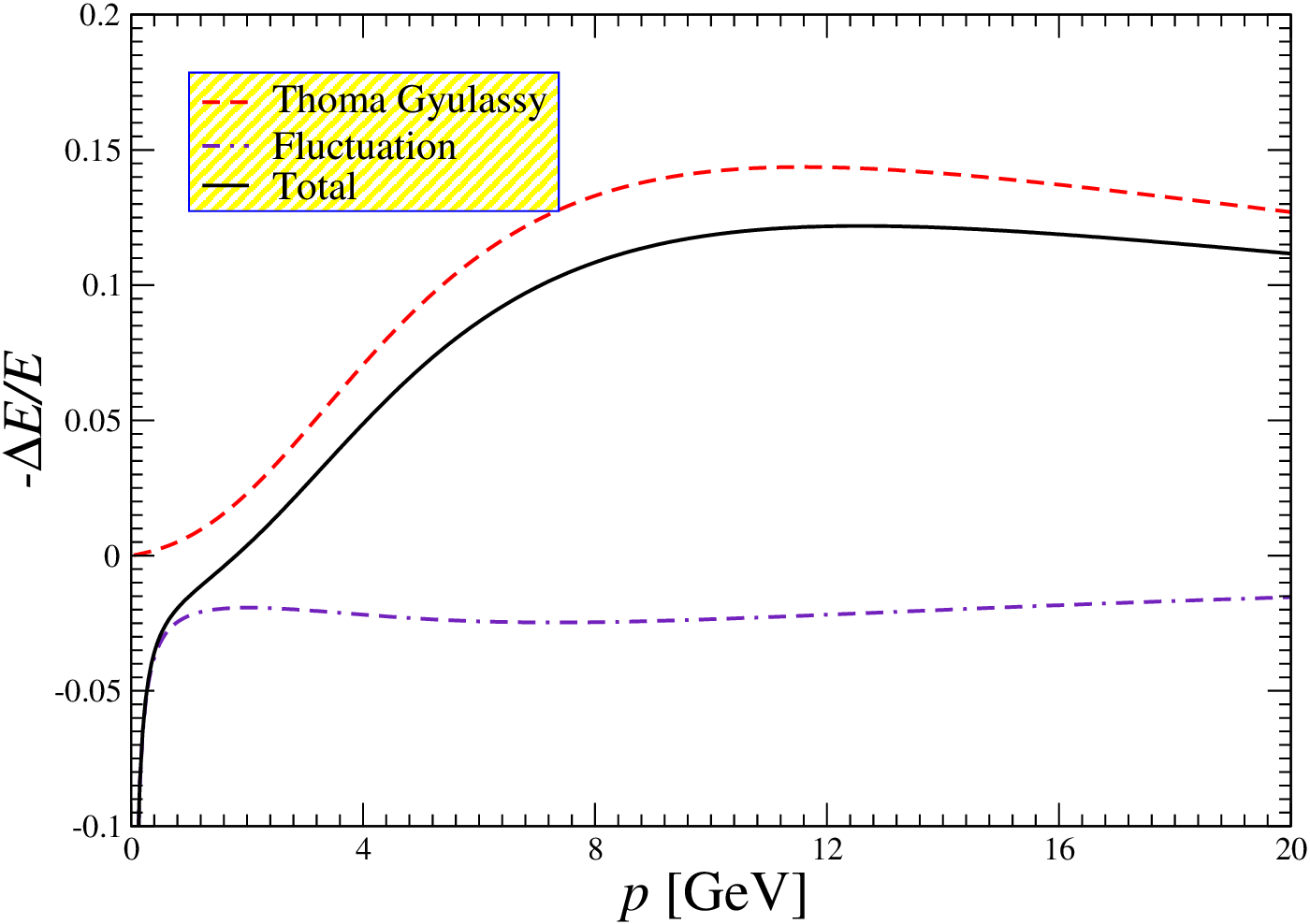}
\caption{\label{fig4} (Color online) Same as Fig.~\ref{fig3} but for 
a bottom quark.} 
\end{center}  
\end{figure} 

In Fig.~\ref{fig3} and Fig.~\ref{fig4} we show the relative collisional 
energy loss of a 
charm and bottom quark where the effect of field fluctuations is taken 
 into account.
It is evident that the effect of the fluctuations on the
heavy quark energy loss is significant at low momenta. For momenta 
$4-20$ GeV the fluctuation effect reduces the collisional loss by 
$17-39\%$ for charm and $12-31\%$ for bottom.
At higher momenta,  as it will be relevant for LHC,
the relative importance of the fluctuation gain to
the collisional loss decreases gradually. The fluctuation gain diverges 
linearly as $v \to 0$. This can be understood by looking at the 
leading log part, 
\begin{eqnarray}
\left.\frac{dE}{dx}\right|_{\rm fl}^{\rm leading-log} &=& 
2\pi C_F\alpha_s^2\left(1+\frac{N_f}{6}\right)\frac{T^3}{Ev^2}
\ln{\frac{1+v}{1-v}} \nonumber \\
&& \times \, \ln{\frac{k_{\rm max}}{k_{\rm min}}}\, 
\label{llog}
\end{eqnarray}
where \( k_{\rm min} = m_D \) is the Debye mass. The divergence here is 
kinematic as $\left.\frac{dE}{dt}\right|_{\rm fl}$ is finite for $v=0$
and therefore $\left.\frac{dE}{dx}\right|_{\rm fl} = \frac{1}{v} \left.\frac{dE}{dt}\right|_{\rm fl}$
diverges for $v \rightarrow 0$.

Let us summarize here the general assumptions made in this investigation:
\begin{itemize}

\item We consider here only the collisional energy loss due to elastic 
scattering.

\item We consider here only heavy quarks (charm, bottom) which became of 
particular interest
in recent experiments at RHIC \cite{Wicks:2005gt}.

\item We use the semiclassical approximation which has been used to calculate 
the mean collisional
energy loss \cite{Thoma:1990fm}. The semiclassical approximation has been 
shown to be equivalent
to the Hard Thermal Loop approximation which is based on the weak coupling 
limit $g \rightarrow 0$. 
It allows a systematic calculation of the collisional energy 
loss~\cite{Braaten:1991} and is valid in the 
high-temperature limit of QCD. It also corresponds to neglecting the non-Abelian terms in the QCD equations 
of motion (see e.g. \cite{thoma1995}).  
Hence we use systematically the Abelian approximation throughout this work. 

\item We assume a constant momentum and temperature independent 
coupling constant. Recently the collisional 
energy loss has been reconsidered using a model to include 
a running temperature dependent coupling constant. Although this leads 
to a different functional dependence on the parameters, the energy loss for realistic situations 
is somewhat larger but of similar size as in the case of a constant coupling \cite{Peshier:2006}.  

\item We assume an equilibrated, isotropic and homogeneous QGP (see below).

\item We assume an infinitely extended QGP. Recently it has been argued that 
finite size effects
are negligible for the collisional energy 
loss~\cite{Djordjevic:2006,Adil:2006ei}. 
  
\end{itemize}  

Let us note here that the assumption of an equilibrium condition necessary 
implies isotropization in momentum space. On the other hand, matter created 
in non-central heavy-ion collisions is anisotropic to start with and the strong 
longitudinal expansion afterwards, at its own, 
brings an anisotropy into the system~\cite{Baym:1984np}. 
The characteristic feature of such anisotropic systems is the presence of 
a Weibel-type instability~\cite{Mrowczynski:1988dz,Mrowczynski:1993qm,
Romatschke:2003ms}. It has been argued that assuming  
a turbulent, weakly coupled anisotropic QGP may provide a 
natural explanation for the observed rapid isotropization time~\cite{Arnold:2004ti}, 
for the small shear viscosity to entropy density 
ratio~\cite{Asakawa:2006jn}, or for dihadron correlation functions~\cite{Romatschke:2006bb}. 
 
The analysis  presented above  can in principle be extended to 
the case of a non-equilibrium, anisotropic QGP, provided the power spectrum of 
the electromagnetic fluctuations are known. One possible way is to simulate 
them on the lattice
\cite{Arnold:2005vb}. Fluctuations are much stronger in non-equilibrium situations
than in thermal systems~\cite{Mrowczynski:1996vh}.  
The effect of field fluctuations on the passage of a heavy quark 
is therefore expected to be stronger in an anisotropic QGP \cite{Note2}.
Interestingly, the energy loss for a heavy quark in an anisotropic QGP  
without taking field fluctuations into account is negative (corresponding
to energy gain) at low energies 
similar to our case~\cite{Romatschke:2004au}. 
Furthermore, in the case of an anisotropic plasma there could be 
an ``anti-Landau'' damping mechanism which would lead to an energy loss 
by fluctuations~\cite{Romatschke:2004au}.

Recently heavy quark probes at RHIC have posed  new challenges to the
theoretical understanding of the parton energy loss. 
As shown by Wicks et al~\cite{Wicks:2005gt}
recent measurement of the non-photonic single electron data cannot 
be explained 
by the radiative loss alone. If the collisional energy loss is included, the 
agreement is better but not satisfactory.  
As seen it is obvious that the gain due to the field fluctuations will lead
to reduction in collisional energy loss and hence more disagreement with single
electron data.
On the other hand, using the transport coefficients within pQCD energy loss
calculations~\cite{Armesto} the elliptic flow coefficient of single electrons 
$v_2$ is limited only to $2-3\%$
in semi-central Au-Au collisions, while the experimental values~\cite{Adler}
reach up to $10\%$ around
transverse electron momenta of $2$ GeV/$c$. It is suggested recently in 
Ref.~\cite{Rapp} that not only the energy loss but also the energy gain 
in low momenta may be required for obtaining larger theoretical 
$v_2$ values.  
It will be interesting to find out whether the inclusion of the 
fluctuation gain can shed light on $v_2$ at low momenta. 

\appendix  

\section{Derivation of (\ref{elossfin1})}
\label{app1}
Since the power spectrum is diagonal in color space according to 
(\ref{eflc}) we can pull 
out the color factor from the electric fields and perform the color sum 
in the second and third terms in (\ref{eqmotion5}) easily. With $Q^a Q^a = C_F
\alpha_s$, where $C_F$ is the quadratic Casimir in fundamental representation, 
we obtain, {\it e.g.}, for the second term in (\ref{eqmotion5})
 \begin{equation}
\label{appeq1}
\# 2 = \frac{C_F\alpha_s}{E_0}\int_0^t\,dt_1\,\left\langle 
\veces\left(\vecr_0\left(t_1\right), t_1\right)\cdot
\veces\left(\vecr_0\left(t\right), t\right)\right\rangle_\beta\,.
\end{equation}
Setting $t-t_1 =\tau$ and provided that there exist scales $\tau_1$ 
and $\tau_2$ discussed earlier, we can write (\ref{appeq1}) as, 
\begin{equation}
\# 2 = \lim_{t\to\infty}
\frac{C_F\alpha_s}{E_0}\int_0^t\,d\tau\, \left\langle \veces\left(\vecr_0
\left(t - \tau\right), t - \tau\right)
\cdot\veces\left(\vecr_0\left(t\right), t\right)\right\rangle_\beta\,.
\label{appeq2}
\end{equation}
Expressing $\veces \left(\vecr_0\left(t\right),t\right)$'s 
in (\ref{appeq2}) in Fourier modes, utilizing the fact that the unperturbed
orbit is a straight line trajectory 
$\vecr_0\left(t\right) = \vecv_0 t$, and  
\begin{eqnarray}
\label{appeq3}
\left\langle \ce_i\left(\vec{k}, \omega\right) 
\ce_j\left(\vec{k}^\prime, \omega^\prime\right)\right\rangle_\beta &=&
\left\langle \ce_i \ce_j \right\rangle_{\beta; \vec{k}, 
 \omega}\left(2\pi\right)^4\delta\left(\omega +\omega^\prime\right)\nonumber \\
&&\delta^3 \left(\vec{k} + \vec{k}^\prime\right)\,,
\end{eqnarray} 
we get, 
\begin{equation}
\label{appeq4}
\# 2 = \frac{C_F\alpha_s}{16\pi^3E_0}\int \,d^3k \left\langle \veces^2
\right\rangle_{\beta; \omega=\vec{k}\cdot\vec{v}_0}
\end{equation} 
Similarly for the third term in (\ref{eqmotion5}) we find, 
\begin{eqnarray}
\label{appeq5}
\# 3 &=& \frac{C_F\alpha_s}{E_0}\int_0^{t}dt_1\int_0^{t_1} dt_2\left\langle
\sum_j\ce_j\left(\vecr_0\left(t_2\right),t_2\right)\right.\nonumber\\
&\times&\left.\frac{\partial}{\partial \vecr_{0j}}\,\,v_{0,i}
\ce_i\left(\vecr_0\left(t\right),t\right)\right\rangle_\beta
\end{eqnarray}
Interchanging the order of integration leads to,
\begin{eqnarray}
\# 3 &=& \frac{C_F\alpha_s}{E_0}\int_0^{t}dt_2\int_{t_2}^{t}dt_1\left\langle
\sum_j\ce_j\left(\vecr_0\left(t_2\right),t_2\right)\right.\nonumber\\
&\times&\left.\frac{\partial}{\partial \vecr_{0j}}\,\,v_{0,i}
\ce_i\left(\vecr_0\left(t\right),t\right)\right\rangle_\beta\,,\nonumber\\
&=& \frac{C_F\alpha_s}{E_0}\int_0^{t}dt_2\left(t - t_2\right)\left\langle
\sum_j\ce_j\left(\vecr_0\left(t_2\right),t_2\right)\right.\nonumber\\
&\times&\left.\frac{\partial}{\partial \vecr_{0j}}\,\,v_{0,i}
\ce_i\left(\vecr_0\left(t\right),t\right)\right\rangle_\beta\,,\nonumber\\
&=&\frac{C_F\alpha_s}{E_0}\lim_{t\to\infty}\int_0^{t}d\tau\,\tau\left\langle
\sum_j\ce_j\left(\vecr_0\left(t-\tau\right),t-\tau\right)\right.\nonumber\\
&\times&\left.\frac{\partial}{\partial \vecr_{0j}}\,\,v_{0,i}
\ce_i\left(\vecr_0\left(t\right),t\right)\right\rangle_\beta\,.\label{appeq6}
\end{eqnarray} 
Using (\ref{appeq3}) and the fact   
the field correlations vanish for $\omega \to \pm \infty $, we get,
\begin{equation}
\label{appeq7}
\# 3 = \frac{C_F\alpha_s}{16\pi^3 E_0}\int\,d^3k\,\,
\omega\frac{\partial}{\partial\omega}
\left\langle \veces_L^2\right\rangle_{\beta; \omega=\vec{k}\cdot\vec{v}_0}
\end{equation}
It is to be noted that in contrast to (\ref{appeq4}) here only the 
longitudinal 
part survives. This is due to the fact that differentiation 
 {\it w.r.t} $r_j$ 
in (\ref{appeq6}) brings down one power of $k_j$ which, operating on field 
correlators (\ref{e.and.b-fluc}), removes the transverse part. 
Adding (\ref{appeq4}) and (\ref{appeq7}) we get (\ref{dedt_fl1}).  

Integrating the first term in (\ref{dedt_fl1}) with respect to the azimuthal angle 
yields, 
\begin{equation}
\label{appeq8}
\# 1 = \frac{C_F\alpha_s}{8\pi^2 E_0}\int_0^{k_{\rm max}}dk\,k^2\int_{-1}
^{+1}d\eta\frac{\partial}{\partial\omega}
\left\langle\omega \veces_L^2\right\rangle_{\beta; \vec{k}\cdot\vec{v}_0}   
\end{equation}
Substituting $\omega = kv_0\eta$ the $\eta$ integration can be performed 
and we get  
\begin{equation}
\label{appeq9}
\# 1 = \frac{C_F\alpha_s}{8\pi^2 E_0 v_0}\int_0^{k_{\rm max}}dk\,k
\left[\left\langle\omega\veces_L^2\right\rangle_
{\omega=kv_0} - \left\langle\omega\veces_L^2\right\rangle
_{\omega=-kv_0}\right]\,.  
\end{equation}
Now $\left\langle\omega \veces_L^2\right\rangle$ can be written as 
$n_B\left(\omega\right)f\left(\omega\right)$ where $n_B\left(\omega\right)$
is the Bose distribution and $f\left(\omega\right) = f\left(-\omega\right)$.
Using $n_B\left(\omega\right) = -\left[1+ n_B\left(\omega\right)\right]$,
and using $\omega = kv_0$ we can write (\ref{appeq9}) as;
\begin{equation}
\label{appeq11}
\# 1 = \frac{C_F\alpha_s}{8\pi^2 E_0v_0^3}\int_0^{k_{\rm max}v_0}d\omega\,
\coth{\frac{\beta\omega}{2}} F\left(\omega, k=\omega /v_0\right)\,.
\end{equation}
Similarly, we obtain the transverse contribution as, 
\begin{equation}
\label{app34}
\# 2 = \frac{C_F\alpha_s}{8\pi^2 E_0v_0}\int_0^{k_{\rm max}}dk\,k\int_0^{kv_0}
d\omega\, \coth{\frac{\beta\omega}{2}} G\left(\omega,k\right)\,.
\end{equation}

\end{document}